\documentclass[12pt]{article}
\usepackage{indentfirst,psfig}
\usepackage{cite}
\usepackage{citesupernumber}
\begin{document}

\title{Monte Carlo Methods for Small Molecule High-Throughput
Experimentation}

\author{
Ligang Chen and Michael W. Deem\\
Department of Chemical Engineering\\
University of California\\
Los Angeles, CA  90095--1592\\
~\\
Corresponding author: M. W. Deem, mwdeem@ucla.edu, 310-267-0169.\\
~\\
To appear in \emph{J.\ Chem.\ Inf.\ Comput.\ Sci.}
}

\maketitle

\begin{abstract}
By analogy with Monte Carlo algorithms, we propose new
strategies for design and redesign of small molecule 
libraries in 
high-throughput experimentation, or
combinatorial chemistry.
 Several Monte Carlo methods are examined, including
Metropolis, three types of biased schemes, and composite moves
that include swapping or parallel tempering. Among them, the biased
Monte Carlo schemes exhibit particularly high efficiency in
locating optimal compounds. The Monte Carlo strategies are compared to
a genetic algorithm approach. Although the best compounds
identified by the genetic algorithm are comparable to those from the
better Monte Carlo schemes, the diversity of favorable compounds
identified is reduced by roughly 60\%.
\end{abstract}
\newpage

\section{Introduction}

High-throughput synthesis is now established as one of the methods for
the discovery of new drugs, materials, and catalysts.
High-throughput, or combinatorial, 
methods allow for simultaneous creation of a large number of
structurally diverse and complex
compounds,\cite{Balkenhohl,Jandeleit,Weinberg} generalizing the
traditional techniques of single compound synthesis. Parallel
synthesis and split/pool synthesis\cite{Furka,wigler} on solid phase,
for example, are two commonly used methods for combinatorial
synthesis. After combinatorial synthesis of the desired number of
compounds, high-throughput screening is used to identify the few
molecules optimally possessing the property of interest. Among the
high-throughput methodologies, small molecule combinatorial chemistry is
the most developed and has been applied successfully in areas such as
transition metal complexation,\cite{Jacobsen} chemical genetic
screening,\cite{Schreiber2} catalysis,\cite{Turner} and drug
discovery.\cite{Schreiber}

High-throughput chemistry can be viewed as a search over a
multi-dimensional space of composition variables for molecules
possessing a high degree of the desired function, or figure of
merit. Indeed, the parallel synthesis and split/pool synthesis methods
search the composition space in a regular, grid-like fashion. As the
complexity of the molecular library grows, the number of dimensions in
the composition variable space grows, and with a grid-like method, the
number of compounds that must be synthesized to search the space grows
exponentially. Synthesis and screening of mixtures of compounds can
partially alleviate the dimensional curse.\cite{Eliseev} However, a
mixture approach raises the question of how to deconvolute and
interpret the results. The greater the degree of mixing, the stronger
the synergistic effects can be in the mixture, and the more difficult
it is to identify individual compounds responsible for the
activity.\cite{Moos} 

The challenge of searching the composition space in an efficient way
has led to extensive efforts in the rational design of combinatorial,
or high-throughput,
libraries. A basic assumption in library design is that structurally
similar compounds tend to display similar activity profiles. By
designing libraries with maximum structural diversity, the potential
for finding active compounds in the high-throughput screenings can be
enhanced. This design approach requires a quantitative account of the
structural and functional diversity of the library, and many
descriptors have been developed.\cite{Ajay} Optimization of a library
to maximum diversity is then driven by a reliable statistical
method. Several structurally diverse libraries have been successfully
designed along these lines.\cite{Bures,Drewry} For example, strategies
have been presented to optimize the structural diversity of libraries
of potential substituents or entire molecules by using stochastic
optimization of diversity functions and a point mutation Monte
Carlo technique.\cite{Hassan} Peptide libraries have been designed by
using topological descriptors and quantitative structure-activity
relationships combined with a genetic algorithm and simulated
annealing.\cite{Tropsha,Tropsha2,Tropsha3} Diverse libraries of
synthetic biodegradable polymers have been designed by using molecular
topology descriptors and a genetic algorithm.\cite{Reynolds} Similarly,
peptoid libraries have been designed by using multivariate
quantitative structure-activity relationships and statistical
experimental design.\cite{Linusson}

The question of how an initial library should be redesigned for
subsequent rounds of high-throughput experimentation in light of the results
of the first round of screening remains unanswered. In this paper, we
suggest that Monte Carlo methods provide a natural means for library
redesign in high-throughput experimentation. The Monte Carlo method is a
well-known statistical method for sampling large spaces efficiently
and ergodically. There is a striking analogy between searching
configuration space for regions of low free energy in a Monte Carlo
simulation and searching composition space for regions of high figure
of merit in a combinatorial chemistry experiment. Importantly, Monte
Carlo methods do not suffer the curse of dimensionality. A Monte Carlo
approach should, therefore, be exponentially more efficient than a
regular, grid-like method for libraries of complex molecules. Indeed,
a Monte Carlo approach to materials discovery proves to
be dramatically more efficient than does a grid-based
approach.\cite{Deem} The application of Monte Carlo methods to small
molecule high-throughput experimentation differs from the conventional
computer simulation technique in several aspects. First, the variables
in molecular high-throughput experimentation are discrete, and no continuous
moves are available. Second, multiple simultaneous searches of the
variable space are performed in high-throughput experimentation
 when screening
the large libraries. Finally, temperature has a natural meaning in
Monte Carlo computer simulation, whereas ``temperature'' is simply a
control parameter in a Monte Carlo protocol for 
high-throughput experimentation.
 In principle, the temperature in the protocol serves to
specify how strong is the differentiation between compounds with low
and high figures of merit. In practice, temperatures that are too low
may cause the method to become trapped in local optima unless a
sufficiently powerful Monte Carlo scheme is used.

Despite the many successes of high-throughput experimentation, the method has
been criticized as a simple machinery, lacking incorporation of {\it a
priori} knowledge when compared with the traditional synthetic
approach. {\it A priori} knowledge, such as chemical intuition,
previous database or experimental information, well-known theory,
patentability, or other specific constraints, are indispensable to an
efficient library design and are the traditional province of the
synthetic chemist. Fascinatingly, the Monte Carlo approach to
high-throughput experimentation can naturally incorporate such knowledge in
the experimental design through the technique of biased Monte Carlo.

Genetic algorithms are the computational analog of Darwinian
evolution. Typically, a genetic algorithm consists of three basic
processes: crossover, mutation, and selection. In the crossover step,
new compounds are generated by mixing the compositions of parent
compounds. In the mutation step, individual molecules are changed at
random. In the selection step, the best molecules are identified for
the next round. The application of genetic algorithms to combinatorial
synthesis and library design has achieved considerable
success.\cite{Gobbi,Weber,Sheridan,Singh}
 Nonetheless, unlike Monte
Carlo algorithms, genetic algorithms do not satisfy detailed
balance. Because of this, genetic algorithms can not be guaranteed to
sample properly the variable space or to locate optimal
molecules. Furthermore, in most experiments, one wants to identify
several initially promising molecules in the hope that, among them, a
few can survive further stringent screenings, such as patentability or
lack of side effects.\cite{Brennan} In the genetic approach, however,
all the molecules in the library tend to become similar to each other
due to the crossover step. 
While diversity can be encouraged in a genetic
approach,\cite{Brown1997,Brown1998} diversity can never 
be guaranteed.
The Monte Carlo approach, on the other
hand, can maintain or even increase the diversity of a molecular
library, due to the satisfaction of detailed balance.

In this paper, we propose several strategies for small molecule
high-throughput experimentation derived by analogy with Monte Carlo
methods. We compare these Monte Carlo protocols to the genetic
algorithm approach. In order to make this comparison and to
demonstrate the effectiveness of the Monte Carlo approach, we perform
simulated high-throughput experiments. In section 2, 
we introduce a random energy
model that we use as a surrogate for experimental measurement of the
figure of merit. 
The random energy model is not fundamental to the protocols;
it is introduced as a simple way to test, parameterize, and
validate the various searching methods.  In an experimental
implementation, the random energy model would be replaced by
the value returned by the screen.
In section 3, we introduce the Monte Carlo protocols
and provide a means to calculate the diversity of a library. In section
4, we compare the various protocols. In section 5, we discuss some
implications of these results. We conclude in section 6.

\section{Space of Variables and A Random Energy Model for Small Molecule 
High-Throughput Experimentation}
\begin{figure}[htbp]
\centering
\leavevmode
\psfig{file=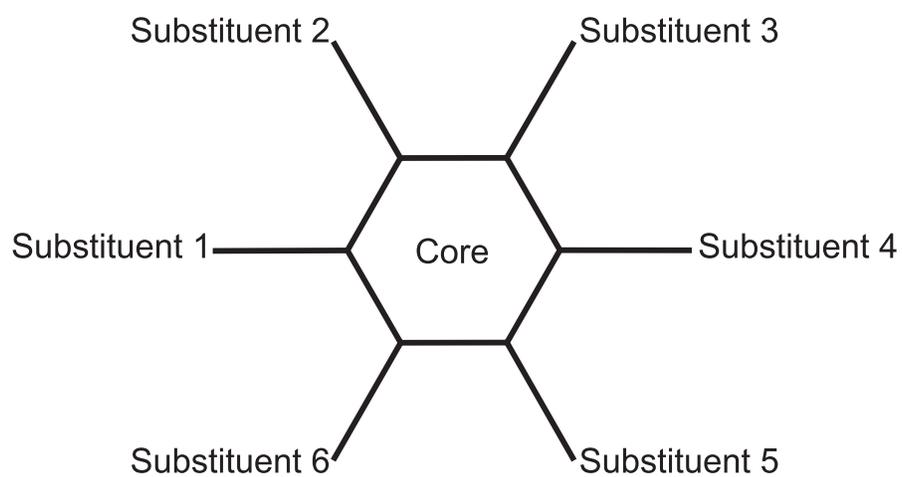,height=2.5in}
\caption{Schematic view of the small molecule model.}
\label{fig1}
\end{figure}

To quantitatively describe the molecules in a high-throughput library,
we uniquely characterize each molecule by its composition, such as the
identity of the core and substituents. For specificity, we will
consider the figure of merit of interest to be a binding constant, but
our results should be generically valid. A schematic view of our model
is presented in Figure \ref{fig1}. For simplicity, we consider the
small molecule to consist of one core, drawn from a library of
cores, and six binding substituents, each drawn from a single library of
substituents.\cite{Katritzky} Numerous energetic interactions could exist
between this molecule and the substrate. It is commonly believed that
descriptors can be directly related to compound performance. A large
class of descriptors, such as one-dimensional, two-dimensional,
three-dimensional, and BCUT descriptors, has been used to measure the
diversity between substituents, cores, and molecules in
the literature.\cite{Ajay,Bures,Drewry} To simplify, we will limit
ourselves to a set of six weakly correlated descriptors for each
substituents and core. For example, the descriptors could be hydrogen
bond donors, hydrogen bond acceptors, flexibility, an
electro-topological calculation, clog$P$, and aromatic
density.\cite{Drewry}

To carry out the simulated experiments, we need a figure of merit
function that mimics the experimental step of measuring the figure of
merit. Once constructed or ``synthesized,'' the molecules are scored
by the model, which takes the composition or molecular descriptors as
input. A random energy model can mimic the generic features of an
experimental figure of merit. For example, the NK model is used to
model combinatorial chemistry experiments on peptides,\cite{Kauffman}
the block NK\cite{Perelson} and generalized NK\cite{Deem2} models are
used to model protein molecular evolution experiments, and the random
phase volume model is used to model materials discovery.\cite{Deem}

The basic building block for our random energy model is a random
polynomial of $n$ descriptors, $x_1, \ldots, x_n$:
\begin{equation}
F(x_1, \ldots, x_n, \{G\}) = \sum_{k=0}^{q}~\sum_{\stackrel{i_1+ \ldots
+i_n=k}{i_1, \ldots, i_n\geq 0}} f_{i_1 \ldots i_n; k}^{\frac{1}{2}}
\xi^{-k}G_{i_1 \ldots i_n}x_1^{i_1} \ldots x_n^{i_n}
\end{equation}
Here $q$ is the degree of the polynomial, $G_{i_1 \ldots i_n}$ are the
fixed coefficients of the polynomial, and $f_{i_1 \ldots i_n; k}$ are
constant symmetry factors,
\begin{equation}
f_{i_1 \ldots i_n; k} = \frac{k!}{i_1! \ldots i_n!}
\end{equation}
We choose to take the square root of $f$ here because we consider each
term of the unsymmetrized polynomial to be random. The scale factor
$\xi$ is used to equalize roughly the contributions from each term of
the polynomial. Since $x_i$ will be drawn from a Gaussian random
distribution of zero mean and unit variance, we set
\begin{equation}
\xi=\left(\frac{\langle x^q \rangle} {\langle x^2
\rangle} \right)^{\frac{1}{q-2}} = \left(\frac{q!}{(q/2)!2^{q/2}}
\right)^{\frac{1}{q-2}}
\end{equation}
We use $q=6$ and $n=6$ in our random energy model. 

The random energy model accounts for contributions to substrate
binding arising from interactions between the substrate and core
and from interactions between the substrate and each of the
substituents. In addition, synergistic effects between the substituents and
core are incorporated. Consider, for example, a molecule made
from core number $m$ from the core library and substituents number
$s_1, \ldots, s_6$ from the substituent library. The core is
characterized by six descriptors, $D_1^{(m)}, \ldots,
D_6^{(m)}$. Similarly, each substituent is characterized by six
descriptors, $d_1^{(s_i)}, \ldots, d_6^{(s_i)}$. We denote the core
contribution to binding by $E_{\rm C}$ and the substituent contributions by
$E_{\rm S}$.  We denote the contribution due to synergistic
substituent-substituent interactions by $E_{\rm SS}$ and the contribution due to
synergistic core-substituent interactions by $E_{\rm CS}$. The total
contribution to the figure of merit is, then,
\begin{equation}
E = E_{\rm S}+E_{\rm C}+E_{\rm SS} + E_{\rm CS}
\end{equation}
Each of these factors is given in terms of the random polynomial:
\begin{eqnarray}
E_{\rm S} &=& \alpha_1 \sum_{i=1}^6 F(d_1^{(s_i)}, \ldots,
d_6^{(s_i)}, \{G_{\rm S}\}) \label{eqn:l}\\ 
E_{\rm C} &=& \alpha_2 F(D_1^{(m)},\ldots, D_6^{(m)},\{G_{\rm C}\})\\
E_{\rm SS} &=& \alpha_3 \sum_{i=1}^{6} h_i F (d_{j_1}^{(s_i)},
d_{j_2}^{(s_i)}, d_{j_3}^{(s_i)}, d_{j_4}^{(s_{i+1})},
d_{j_5}^{(s_{i+1})}, d_{j_6}^{(s_{i+1})}, \{G_{\rm SS}\})
\label{eqn:ll}\\ 
E_{\rm CS} &=& \alpha_4 \sum_{i=1}^{6} h_i F(d_{k_1}^{(s_i)},
d_{k_2}^{(s_i)}, d_{k_3}^{(s_i)}, D_{k_4}^{(m)}, D_{k_5}^{(m)},
D_{k_6}^{(m)}, \{G_{\rm CS}\})
\label{eqn:lt}
\end{eqnarray}
where the $\{G_{\rm S}\}, \{G_{\rm C}\}, \{G_{\rm SS}\}$, and
$\{G_{\rm CS}\}$ are four sets of fixed random Gaussian variables with
zero mean and unit variance. The $\alpha_i$ are constants to be
adjusted so that the synergistic terms will contribute in desired
percentages, and $h_i$ is a structural constant indicating the
strength of the interaction at binding site $i$. The interaction
strengths $h_i$ are chosen from a Gaussian distribution of zero mean
and unit variance for each site on each core. Only synergistic
interactions between neighboring substituents are considered in $E_{\rm
SS}$, and it is understood that $s_7$ refers to $s_1$ in
eq~\ref{eqn:ll}. In principle, the polynomial in eq~\ref{eqn:ll}
could be a function of all 12 descriptors of both substituents. We assume,
however, that important contributions come from interactions among
three randomly chosen distinct descriptors of substituent $s_i,
d_{j_1}^{(s_i)}, d_{j_2}^{(s_i)},$ and $d_{j_3}^{(s_i)},$ and another
three randomly chosen distinct descriptors of substituent $s_{i+1},
d_{j_4}^{(s_{i+1})}, d_{j_5}^{(s_{i+1})},$ and
$d_{j_6}^{(s_{i+1})}$. Similarly, we assume that core-substituent
contributions come from interactions between three randomly chosen
distinct descriptors of the substituent, $d_{k_1}^{(s_i)},
d_{k_2}^{(s_i)},$ and $d_{k_3}^{(s_i)},$ and another three randomly
chosen distinct descriptors of the core, $D_{k_4}^{(m)},
D_{k_5}^{(m)},$ and $D_{k_6}^{(m)}$. Both $j_i$ and $k_i$ are
descriptor indices ranging from 1 to 6. Assuming that we have
integrated out the degrees of freedom of the substrate, these indices
depend only on the core.

The parameters in the random energy model are chosen to mimic the
complicated interactions between a small molecule
and a substrate.  We choose to focus on the case where
these interactions are unpredictable, which is typical.
That is, in a typical experiment, it would not be possible
to predict the value of the screen in terms of molecular descriptors.
Indeed, when rational design fails, an intelligent use of
high-throughput experimentation is called for.
The task of library design and redesign, rather than single molecule
design, is the one we address in the next section.

\section{Monte Carlo Strategies and Diversity Measurement}
Before initiating the Monte Carlo protocol, we first build the
core and substituent libraries. We denote the size of the core
library by $N_{\rm C}$ and the size of the substituent library by $N_{\rm
S}$. In a real experiment, the six descriptors would then be
calculated for each core and substituent. In the simulated experiment,
the values of the six descriptors of each substituent and core are
extracted from a Gaussian random distribution with zero mean and unit
variance. In the simulated experiment, we also associate two sets of
random interaction descriptor indices to each core for the
interaction terms in eqs~\ref{eqn:ll} and~\ref{eqn:lt}.

To give a baseline for comparison, we first design the library
using a random construction.
  New molecules are constructed by random selection of one
core and six substituents from the libraries. Since the properties of
each substituent and core are assigned randomly, this first library
should be reasonably diverse and comparable to examples in literature.

For the Monte Carlo schemes, the initial molecular configurations are
assigned randomly as before. The library is modified by the Monte
Carlo protocol in subsequent rounds of high-throughput experimentation. Two
kinds of move are possible for each molecule in the library, core
changes and substituent changes. Either the core is changed with
probability $p_{\rm core}$, or one of the six substituents is picked
randomly to change. We denote the probability of changing from
core $m$ to $m'$ by $T(m \rightarrow m')$ and from substituent $i$ to
$i'$ by $t(i \rightarrow i')$. The new configurations are updated
according to the acceptance rule at $\beta$, the inverse of the
protocol temperature. All the samples are sequentially updated in one
Monte Carlo round.

For the simple Metropolis method, the transition matrices are
\begin{eqnarray}
T(m \rightarrow m') &=& 1/N_{\rm C} \\
t(i \rightarrow i') &=& 1/N_{\rm S}
\end{eqnarray}
and the acceptance rule is
\begin{equation}
{\rm acc}(o \rightarrow n)=\min[1,~\exp(-\beta \Delta E)]
\end{equation}
To make use of the idea that smaller moves are accepted more often, we
could try to choose a modified substituent or core that is similar to
current one, that is, we could use a transition matrix weighted
towards those substituents or cores close to the current one in the
six-dimensional descriptor space. Interestingly, this refinement turns
out not to work any better than does the simple random move. It seems
that even a small move in the descriptor space is already much larger
than the typical distance between peaks on the figure of merit
landscape.

Biased Monte Carlo methods have been shown to improve the sampling of
complex molecular systems by many orders of magnitude.\cite{Frenkel}
In contrast to conventional Metropolis Monte Carlo, trial moves in
biased schemes are no longer chosen completely at random. By
generating trial configurations with a probability that depends on
{\it a priori} knowledge, the moves are more likely to be favorable
and more likely to be accepted. As we are dealing with a discrete
configurational space, the implementation of biased Monte Carlo in
this case is relatively simple. First, we need a biasing term for both
substituents and cores. Since the form of this term is not unique, we
can proceed in several different ways. One strategy is to bias our
choice of core and substituent on the individual contributions of the
cores and substituents to the figure of merit. We might know, or be
able to estimate, these contributions from theory. For the random
energy model, for example, we know
\begin{eqnarray}
e^{(i)} &=& \alpha_1 F(d_1^{(i)}, \ldots, d_6^{(i)}, \{G_{\rm S}\})\\
E^{(m)} &=& \alpha_2 F(D_1^{(m)},\ldots, D_6^{(m)},\{G_{\rm C}\})
\end{eqnarray}
where $e^{(i)}$ is the bias energy to the substituent $i$ in the library,
and $E^{(m)}$ the bias energy of core $m$ in the library.
Alternatively, we can estimate the contribution of each substituent or
core to the figure of merit
experimentally.\cite{Fesik,Moore,Ellman} A electrospray ionization
source coupled to a mass spectrometer, for example, can serve this
purpose.\cite{Griffey} To measure the contributions, we do a
pre-experiment on 10000 randomly constructed molecules. This number of
compounds will give on average each substituent 60 hits and each core
667 hits. By averaging the figure of merit of the molecules containing
a particular substituent or core over the total number of hits, we can
obtain experimental estimates of $e^{(i)}$ and $E^{(m)}$. Using these
two methods of bias, we construct three different types of biased
Monte Carlo schemes: theoretical biased move, experimental biased
move, and mixed biased move. In theoretical bias, both $e^{(i)}$ and
$E^{(m)}$ are from the random energy model. In experimental bias, both
$e^{(i)}$ and $E^{(m)}$ are calculated from the pre-experiment. In
mixed bias, $e^{(i)}$ comes from the random energy model, while
$E^{(m)}$ comes from the pre-experiment.

These biases tend to exhibit a large gap between a few dominant
cores and substituents and the rest. To ensure the participation of
more substituents and cores in the strategy, we introduce cutoff
energies for the substituent and core, $e_{\rm c}$ and $E_{\rm c}$.  We
arbitrarily choose $e_{\rm c}$ to be the 21st lowest substituent energy and
$E_{\rm c}$ to be the 4th lowest core energy.  The biased energy,
$e_{\rm b}^{(i)}$, for the $i^{th}$ substituent is
\begin{equation}
e_{\rm b}^{(i)}=\left\{ \begin{array}{ll}
		e^{(i)} & \mbox{if $e^{(i)}>e_{\rm c}$} \\
		e_{\rm c} & \mbox{otherwise}
		\end{array}
	\right.
\end{equation}
And the biased energy, $E_{\rm b}^{(m)}$, for the $m^{th}$ core is
\begin{equation}
E_{\rm b}^{(m)}=\left\{ \begin{array}{ll}
		E^{(m)} & \mbox{if $E^{(m)}>E_{\rm c}$} \\
		E_{\rm c} & \mbox{otherwise}
		\end{array}
	\right.
\end{equation}

To correct for this bias, we introduce Rosenbluth
factors.\cite{Frenkel} Since the transition probabilities are the same
at each Monte Carlo step, we have a constant Rosenbluth factor for the
substituent
\begin{equation}
w(n)=w(o)=\sum_{i=1}^{N_{\rm S}} \exp(-\beta e_{\rm b}^{(i)})
\end{equation}
The probability of transition from substituent $i$ to $i'$ is
\begin{equation}
t(i \rightarrow i')=\frac{\exp(-\beta e_{\rm b}^{(i')})}{w(n)}
\end{equation}
In the same way, the Rosenbluth factor for the core is
\begin{equation}
W(n)=W(o)=\sum_{m=1}^{N_{\rm C}} \exp(-\beta E_{\rm b}^{(m)})
\end{equation}
The probability of transition from core $m$ to $m'$ is
\begin{equation}
T(m \rightarrow m')=\frac{\exp(-\beta E_{\rm b}^{(m')})}{W(n)}
\end{equation}
Finally, we define the remaining, non-biased part of the figure of
merit to be
\begin{equation}
E_{\rm b}=E-E_{\rm b}^{(m)}-\sum_{i=1}^6 e_{\rm b}^{(s_i)}
\end{equation}
To satisfy the detail balance, the acceptance rule becomes
\begin{equation}
{\rm acc}(o \rightarrow n) = \min[1, \exp(-\beta \Delta E_{\rm b})]
\end{equation}

We add a swap move that attempts to exchange fragments between two
molecules to the set of Monte Carlo moves. We denote the probability
of attempting a swap instead of a single-molecule move as $p_{\rm
swap}$. In a swap move, the cores or a pair of substituents may be
swapped between two randomly selected molecules. The probability of
switching the core or substituent at the same position is given by
$p_{\rm swap_C}$ and $p_{\rm swap_S}$, respectively. The crossover
event from genetic algorithms could also be introduced in the swap
moves, but this additional move did not improve the results. The
acceptance rule for swapping is ${\rm acc}(o \rightarrow
n)=\min[1,~\exp(-\beta \Delta E)]$.

Parallel tempering is known to be a powerful tool for searching rugged
energy landscapes.\cite{Swendsen,Geyer} In parallel tempering, the
samples are divided into $k$ groups. The first group of samples is
simulated at $\beta_1$, the second group is at $\beta_2$, and so on,
with $\beta_1< \beta_2 <\ldots < \beta_k$. At the end of each round,
samples in group $i<k$ are allowed to exchange configurations with
samples in group $i+1$ with probability $p_{\rm ex}$. The
corresponding acceptance rule for a parallel tempering exchange is
\begin{equation}
{\rm acc}(o \rightarrow n)=\min[1,~\exp(\Delta \beta~\Delta E)]
\end{equation}
where $\Delta\beta=\beta_i-\beta_{i+1}$ and $\Delta E$ is the
difference in energy between the sample in group $i$ and the sample in
group $i+1$. It is important to notice that this exchange step is
experimentally cost-free. Nonetheless, this step can be dramatically
effective at facilitating the protocol to escape from local
minimum. The number of groups, the number of samples in each group,
the value of $\beta_i$, and the exchange probability, $p_{\rm ex}$, are
experimental parameters to be tuned.

For comparison, we compare these Monte Carlo protocols to a standard
genetic algorithm approach.\cite{Gobbi,Weber,Sheridan,Singh} In the
genetic algorithm, as in the Monte Carlo strategy, we perform multiple
rounds of experimentation on a large set of compounds. The difference
between the Monte Carlo and the genetic algorithm lies in how the
library is redesigned, that is, how the compounds are modified in each
round. In the genetic algorithm, first we randomly select two
parents. Then we list the explicit composition of each molecule,
i.e. core, substituent 1, \ldots, substituent 6. After aligning the
sequences from the two parents, we make a random cut and exchange part
of the sequences before the cut. We also allow for random changes, or
mutation, in the cores or substituents of the offsprings. Finally, since
the population is doubled by crossover, we select the better half of
the molecules to survive this procedure and continue on to the next
round.

The diversity of the library as it passes through the rounds of
high-throughput experimentation is an important quantity. We calculate the
diversity, ${\cal{D}}$, as the standard deviation of
the library in the 42-dimensional descriptor space.
\begin{equation}
{\cal{D}}^2 = \frac{1}{N} \sum_{i=1}^N \left[\sum_{j=1}^6
(D_j^{(m(i))}-\langle D_j \rangle)^2 + \sum_{j=1}^6 \sum_{k=1}^6
(d_j^{(s_k(i))} - \langle d_j^{(s_k)} \rangle)^2 \right]
\label{eqn:div}
\end{equation}
where $m(i)$ is the index of the core of molecule $i$, $s_k(i)$ is
the index of substituent $k$ of molecule $i$, and $j$ is the index for the
descriptor. The average value in each descriptor dimension is given by
$\langle D_j \rangle = N^{-1} \sum_{i=1}^{N} D_j^{(m(i))}$ and
$\langle d_j^{(s_k)}\rangle = N^{-1} \sum_{i=1}^{N}
d_j^{(s_k(i))}$. The diversity of the library will change as the
library changes. A larger library will generally possess a higher
absolute diversity simply due to the increased number of compounds. This
important, but trivial, contribution to the diversity is scaled out by
the factor of $1/N$ in eq~\ref{eqn:div}.

\section{Results}

To gauge how the synergistic terms in the figure of merit affect the
efficiency of the Monte Carlo protocols, we consider three models with
increasingly important synergistic effects. We do this by adjusting
the $\alpha_i$ in eqs~\ref{eqn:l}--\ref{eqn:lt}, so that the absolute
values of the terms are on average in the ratio $E_{\rm S}:E_{\rm
C}:E_{\rm SS}:E_{\rm CS}=1:1:0.5:0.3$ in model I, $1:1:1:0.6$ in model
II, and $1:1:2:1.2$ in model III. Finally, we set $\alpha_1=0.01$
arbitrarily in model I. To maintain the same statistical magnitude of
total energy, we set $\alpha_1=0.00778$ in model II and
$\alpha_1=0.00538$ in model III.

The size of the library is fixed at $N_{\rm C}=15$ and $N_{\rm
S}=1000$. The compositional space of this model has $15\times 1000^6$
distinct molecules. Clearly, it is impossible to search exhaustively
even this modestly complex space. We fix the total number of molecules
to be synthesized at 100000, that is, all protocols will have roughly
the same experimental cost. Specifically, 100000 molecules will be
made in the random library design protocol, while in the case of the
Monte Carlo or genetic protocols, the number of molecules times the
number of simulation rounds is kept fixed at 100000.

To locate optimal parameters for the protocols, we perform a few short
pre-experiments. We first fix the energy coefficients in the energy
function and the descriptors of the substituent and core libraries.
For simple Metropolis, we find that it is optimal to use 10 samples
with 10000 rounds, suggesting that the system is still far from
equilibrium at the random initial configuration. With the biased Monte
Carlo method, we find that 100 samples and 1000 rounds is optimal.  We
focus on systems with 1000 or 100 rounds, since fewer rounds are
typically preferred in experiments. It is more difficult to achieve
effective sampling in the system with 100 rounds, and so we use
this system when setting optimal parameter values. For parallel tempering,
it was optimal to have the samples divided into three subsets, with
30\% of the population at $\beta_1$, 40\% at $\beta_2$, and 30\% at
$\beta_3$. The optimal parameters are listed in Table I for each
model. Determination of these parameter values corresponds
experimentally to gaining familiarity with the protocol on a new
system.

\begin{figure}[htbp]
\centering
\leavevmode
\psfig{file=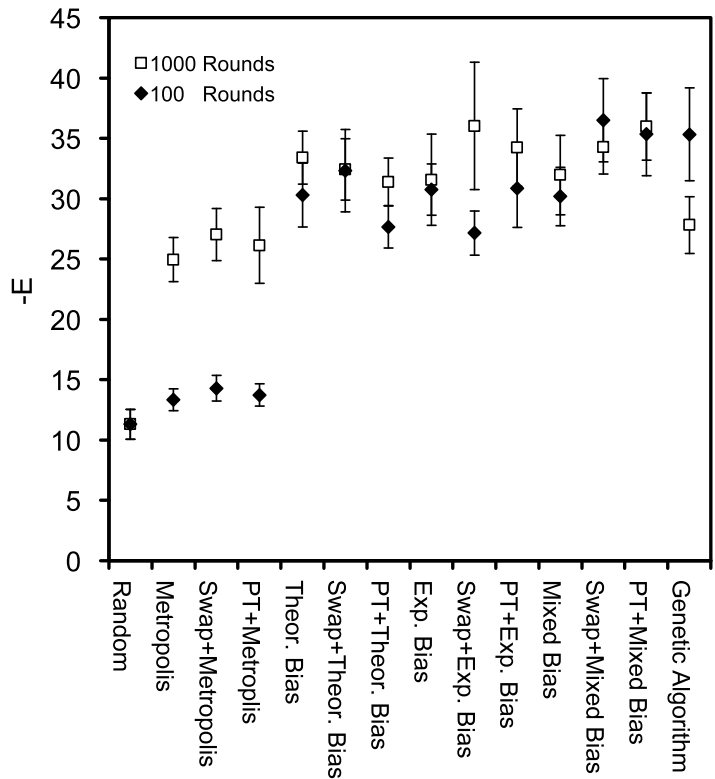,height=5in}
\caption{Comparison of different Monte Carlo schemes with random and
genetic schemes for energy model I ($E_{\rm S}:E_{\rm C}:E_{\rm
SS}:E_{\rm CS}=1:1:0.5:0.3$). Data from two cases are shown, one with
1000 molecules and 100 rounds (filled diamonds) and one with 100
molecules and 1000 rounds (unfilled squares).
Only comparison between relative energy values is meaningful,
as the energy scale is abritrary.
}
\label{model1}
\end{figure}

\begin{figure}[htbp]
\centering
\leavevmode
\psfig{file=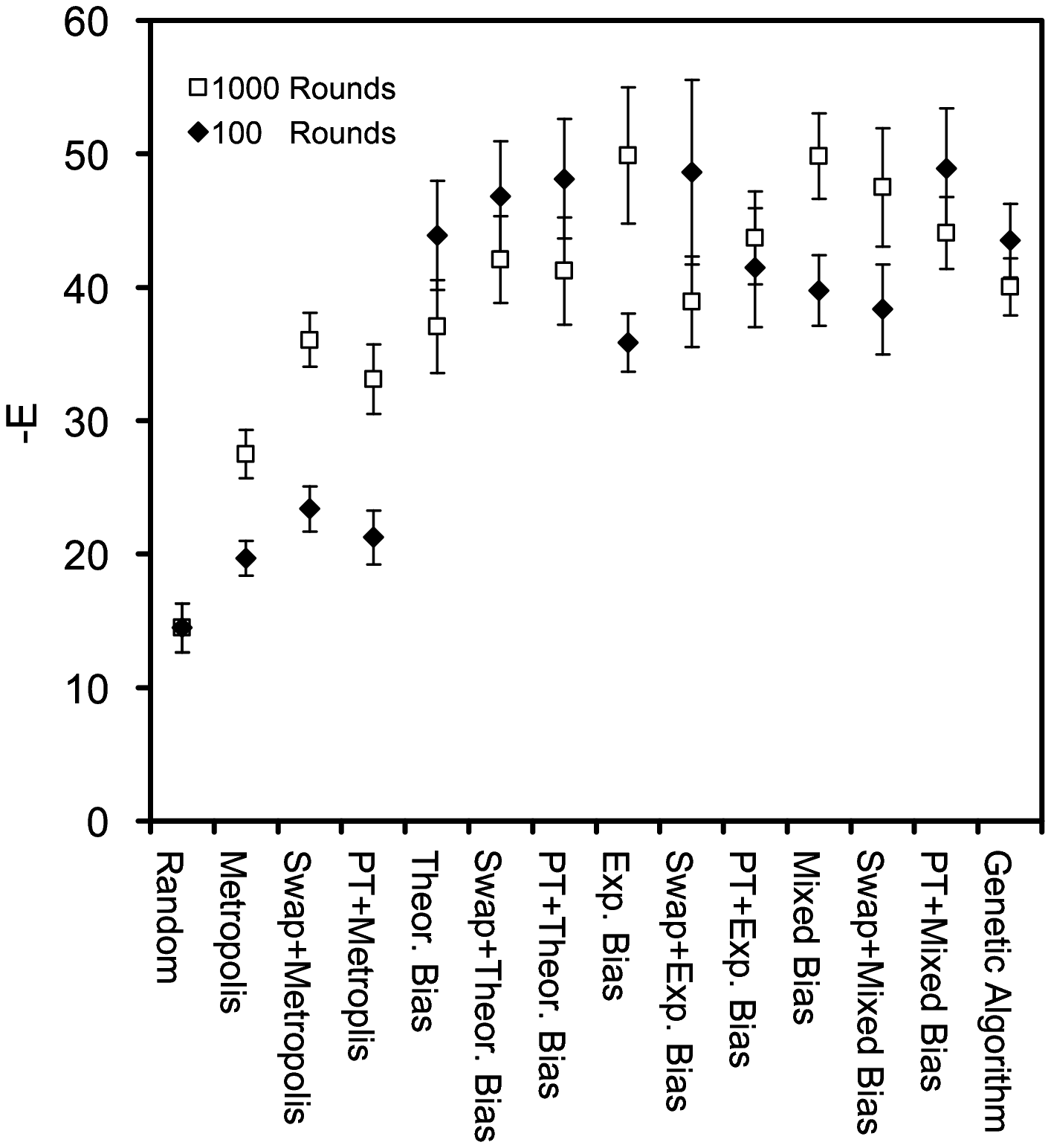,height=5in}
\caption{Comparison of different Monte Carlo schemes with random and
genetic schemes for energy model II ($E_{\rm S}:E_{\rm C}:E_{\rm
SS}:E_{\rm CS}=1:1:1:0.6$). Data from two cases are shown, one with
1000 molecules and 100 rounds (filled diamonds) and one with 100
molecules and 1000 rounds (unfilled squares).}
\label{model2}
\end{figure}

\begin{figure}[htbp]
\centering
\leavevmode
\psfig{file=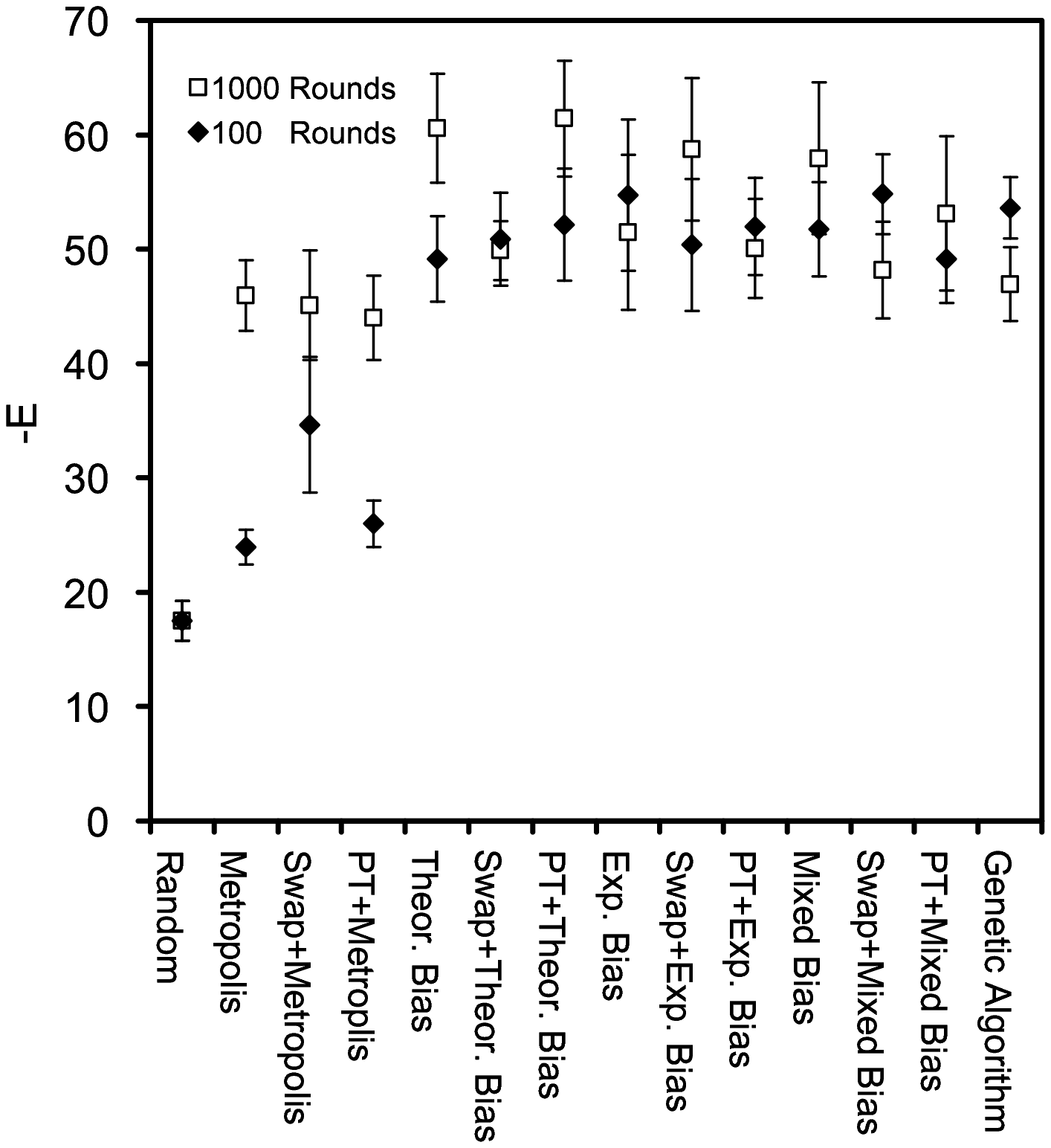,height=5in}
\caption{Comparison of different Monte Carlo schemes with random and
genetic schemes for energy model III ($E_{\rm S}:E_{\rm C}:E_{\rm
SS}:E_{\rm CS}=1:1:2:1.2$). Data from two cases are shown, one with
1000 molecules and 100 rounds (filled diamonds) and one with 100
molecules and 1000 rounds (unfilled squares).}
\label{model3}
\end{figure}

The various Monte Carlo schemes are compared with the random selection
method and the genetic algorithm. Once the optimal parameters are
chosen, the coefficients of the energy function and the descriptor
values of the substituent and core libraries are generated differently
in each simulated experiment. The simulation results for the three
models are shown in Figures~\ref{model1}--\ref{model3}. Each data point
in the figure is an average over 20 independent runs. This averaging
is intended to give representative performance of the protocols on
various figures of merit of experimental interest. Since there is much
randomness in the results, the standard deviation of the average is
shown as well.

\section{Discussion}
Although the average absolute values of the figures of merit in the
three models are adjusted to be equal, the stronger the interaction
terms, the greater the figure of merit we can find. For instance,
the biased schemes find values of $-E$ in the range 30--40 in model I, but
values in the range 40--50 in model II and values in the range 50--60 in
model III. This suggests that the figure of merit landscape has
changed in detail as the synergistic effects in the model are
adjusted. It is clear that for all systems, the Metropolis methods
perform better than does random selection. The system with 1000
molecules and 100 rounds is not well-equilibrated by the Metropolis
schemes, and an experiment with 100 molecules and 1000 rounds
significantly improves the optimal compounds identified. However, by
incorporating {\it a priori} knowledge, the biased Monte Carlo schemes
are able to equilibrate the experiment with either 1000 or 100
rounds. Interestingly, the theoretical bias and experimental bias
methods yield similar results. This strongly suggests that a minimal
number of pre-experiments can be very useful, both for the
understanding of the structure of the figure of merit landscape and
for improving the performance in future rounds.

The results produced with the composite moves including swap and
parallel tempering are slightly improved relative to those from the
plain Monte Carlo schemes. Typically, however, these composite moves
significantly improve the sampling of a rough landscape. Indeed,
swapping and crossover moves are very effective in protein molecular
evolution, where the variable space is extremely large.\cite{Deem2}
Perhaps the variable space is not so large in small molecule
high-throughput experimentation that these composite moves are
required. Alternatively, the random energy model may underestimate the
ruggedness of the landscape. The landscape for RNA substituents, for
example, is estimated to be extremely rough,\cite{Griffey} and
composite moves may prove more important in this case.

\begin{figure}[htbp]
\centering
\leavevmode
\psfig{file=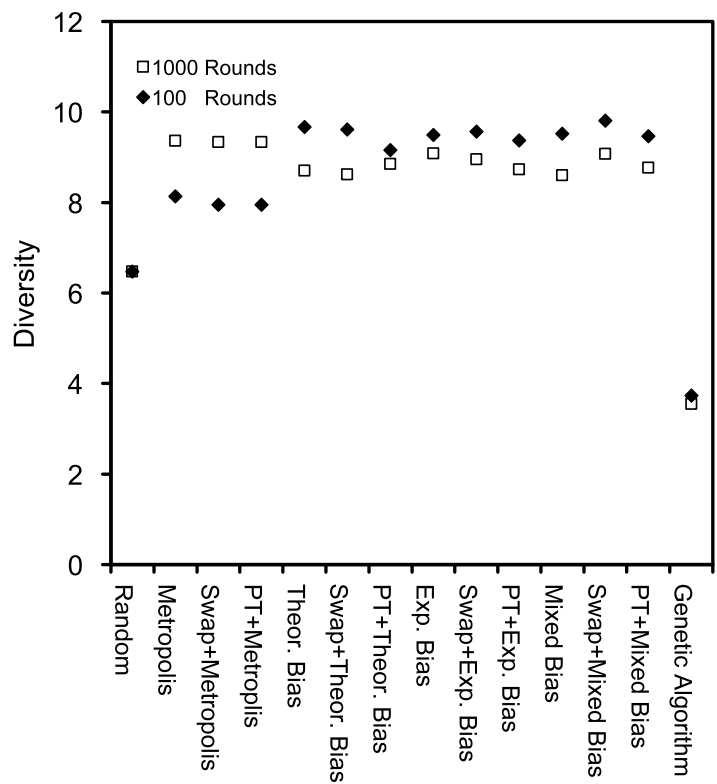,height=5in}
\caption{Diversity measurement of the final configurations for model
I. Data from two cases are shown, one with 1000 molecules and 100
rounds (filled diamonds) and one with 100 molecules and 1000 rounds
(unfilled squares). The error bars are negligible. The contribution to
the absolute diversity that scales as the square root of the number of
molecules per round has been scaled out in this figure, as in
eq~\ref{eqn:div}.}
\label{diversity}
\end{figure}

The genetic algorithm is relatively easy to use. It does not satisfy
detailed balance, however, so there is no theoretical guarantee of the
outcome. The optimal figures of merit identified are, nonetheless,
comparable to those from the better Monte Carlo methods for all three
models. However, due to the crossover and selection steps in the
genetic algorithm, the molecules in the library tend to become similar
to each other, which prevents this scheme from sampling the whole
variable space. To help elucidate this point, diversity measurements
for model I are shown in Figure~\ref{diversity}. It is clear that the
genetic algorithm has reduced the diversity of the library by 60\%
relative to the biased schemes. Interestingly, the Monte Carlo
simulations actually increase the diversity from the initial random
configurations.  The biased schemes tend to bring the system to
equilibrium relatively quickly, and the diversity measurements are
similar for the 100 and 1000 round experiments. For the Metropolis
method, on the other hand, an experiment with 100 rounds is less
diverse than an experiment with 1000 rounds. The genetic algorithm
approach finds less favorable figure of merit values in the 100
compound 1000 round experiment, presumably due to a greater
sensitivity to the $\sqrt{10}$ reduction in the absolute diversity
relative to the already small absolute diversity in the 1000 compound
100 round experiment.

The greater the number of potentially favorable molecules in the
library space, the greater the diversity of the experimental library
will be for the Monte Carlo methods. The genetic algorithm, on the
other hand, will tend to produce a library that contains many copies
of a single favorable molecule. A key distinction, then, is that a
Monte Carlo strategy will sample many compounds from the figure of
merit landscape, whereas a genetic algorithm will tend to produce a
single molecule with a favorable figure of merit value. How strongly
the compounds with high figures of merit are favored in the Monte
Carlo strategy is determined by the protocol temperature, since the
probability of observing a compound with figure of merit $-E$ is
proportional to $\exp(-\beta E)$. The sampling achieved by the Monte
Carlo methods is important not only because it assures that the
composition space is thoroughly sampled, but also because it assures
that the library of final hits will be as diverse as possible.

The Monte Carlo methods perform equally well on all three models.  The
three models were introduced to gauge the impact of unpredictable,
synergistic effects in the experimental figure of merit. It might be
expected that the {\it a priori} bias methods would perform less well
as the synergistic effects become more pronounced. That the biased
methods perform well even in model III suggests that the Monte Carlo
approach may be rather robust. In other words, even a limited amount
of \emph{a priori} information is useful in the Monte Carlo approach
to library redesign.

\section{Conclusion}
Monte Carlo appears to be a fruitful paradigm for experimental design
of multi-round combinatorial chemistry, or
high-throughput, experiments. A criticism of
high-throughput experimentation has been its mechanical structure and lack of
incorporation of {\it a priori} knowledge. As shown here, a biased
Monte Carlo approach handily allows the incorporation of {\it a
priori} knowledge. Indeed, our simulation results reveal that biased
Monte Carlo schemes greatly improve the chances of locating optimal
compounds. For the moderately complex libraries considered here, the
bias can be determined equally well by experimental or theoretical
means. Although the compounds identified from a traditional genetic
algorithm are comparable to those from the better Monte Carlo schemes,
the diversity of identified molecules is dramatically decreased in the
genetic approach. Genetic algorithms, therefore, are less suitable
when the list of good molecules is further winnowed by a secondary
screen, a tertiary screen, patentability considerations, lack of side
effects, or other concerns. Interestingly, composite Monte Carlo
moves such as swap or parallel tempering bring only a slight
improvement to the plain biased Monte Carlo protocols, possibly due to
the relatively small size of the composition space in small molecule
high-throughput experimentation. Presumably, as the complexity of the library
is increased, these composite moves will prove more useful for the
more challenging figures of merit. Although we have here chosen the
initial library configurations at random, the sophisticated initial
library design strategies available in the literature can be used, and
they would complement the multi-round library redesign strategies
presented here.
\section*{Acknowledgments}
This research was supported by National Science Foundation through
grant number CTS--9702403.

\bibliography{cchem}

\providecommand{\refin}[1]{\\ \textbf{Referenced in:} #1}
\begin{thebibliography}{10}

\bibitem{Balkenhohl}
Balkenhohl,~F.;\ \ von~dem Bussche-H\"{u}nnefeld,~C.;\ \ Lansky,~A.;\ \
  Zechel,~C. \textit{Angew. Chem. Int. Ed. Engl.} \textbf{1996,} \textsl{35,}
  2288-2377.

\bibitem{Jandeleit}
Jandeleit,~B.;\ \ Schaefer,~D.~J.;\ \ Powers,~T.~S.;\ \ Turner,~H.~W.;\ \
  Weinberg,~W.~H. \textit{Angew. Chem. Int. Ed.} \textbf{1999,} \textsl{38,}
  2494-2532.

\bibitem{Weinberg}
McFarland,~E.~W.;\ \ Weinberg,~W.~H. \textit{Trends Biotechnol.} \textbf{1999,}
  \textsl{17,} 107-115.

\bibitem{Furka}
Furka,~A.;\ \ Sebestyen,~F.;\ \ Asgedom,~M.;\ \ Dibo,~G. \textit{Int. J.
  Peptide Protein Res.} \textbf{1991,} \textsl{37,} 487-493.

\bibitem{wigler}
Ohlmeyer,~M. H.~J.;\ \ Swanson,~R.~N.;\ \ Dillard,~L.~W.;\ \ Reader,~J.~C.;\ \
  Asouline,~G.;\ \ Kobayashi,~R.;\ \ Wigler,~M.;\ \ Still,~W.~C. \textit{Proc.
  Natl. Acad. Sci. USA} \textbf{1993,} \textsl{90,} 10922-10926.

\bibitem{Jacobsen}
Francis,~M.~B.;\ \ Jamison,~T.~F.;\ \ Jacobsen,~E.~N. \textit{Curr. Opin. Chem.
  Biol.} \textbf{1998,} \textsl{2,} 422-428.

\bibitem{Schreiber2}
Tan,~D.~S.;\ \ Foley,~M.~A.;\ \ Stockwell,~B.~R.;\ \ Shair,~M.~D.;\ \
  Schreiber,~S.~L. \textit{J. Am. Chem. Soc.} \textbf{1999,} \textsl{121,}
  9073-9087.

\bibitem{Turner}
Weinberg,~W.~H.;\ \ Jandeleit,~B.;\ \ Self,~K.;\ \ Turner,~H. \textit{Curr.
  Opin. Solid State Mat. Sci.} \textbf{1998,} \textsl{3,} 104-110.

\bibitem{Schreiber}
Schreiber,~S.~L. \textit{Science} \textbf{2000,} \textsl{287,} 1964-1969.

\bibitem{Eliseev}
Nazarpack-Kandlousy,~N.;\ \ Zweigenbaum,~J.;\ \ Henion,~J.;\ \ Eliseev,~A.~V.
  \textit{J. Comb. Chem.} \textbf{1999,} \textsl{1,} 199-206.

\bibitem{Moos}
Desai,~M.~C.;\ \ Zuckermann,~R.~N.;\ \ Moos,~W.~H. \textit{Drug Develop. Res.}
  \textbf{1994,} \textsl{33,} 174-188.

\bibitem{Ajay}
Ajay,;\ \ Walters,~W.~P.;\ \ Murcko,~M.~A. \textit{J. Med. Chem.}
  \textbf{1998,} \textsl{41,} 3314-3324.

\bibitem{Bures}
Bures,~M.~G.;\ \ Martin,~Y.~C. \textit{Curr. Opin. Chem. Biol.} \textbf{1998,}
  \textsl{2,} 376-380.

\bibitem{Drewry}
Drewry,~D.~H.;\ \ Young,~S.~S. \textit{Chemometr. Intell. Lab.} \textbf{1999,}
  \textsl{48,} 1-20.

\bibitem{Hassan}
Hassan,~M.;\ \ Bielawski,~J.~P.;\ \ Hempel,~J.~C.;\ \ Waldman,~M. \textit{Mol.
  Divers.} \textbf{1996,} \textsl{2,} 64-74.

\bibitem{Tropsha}
Zheng,~W.;\ \ Cho,~S.~J.;\ \ Tropsha,~A. \textit{J. Chem. Inf. Comp. Sci.}
  \textbf{1998,} \textsl{38,} 251-258.

\bibitem{Tropsha2}
Zheng,~W.;\ \ Cho,~S.~J.;\ \ Tropsha,~A. \textit{J. Chem. Inf. Comp. Sci.}
  \textbf{1998,} \textsl{38,} 259-268.

\bibitem{Tropsha3}
Zheng,~W.;\ \ Cho,~S.~J.;\ \ Waller,~C.~L.;\ \ Tropsha,~A. \textit{J. Chem.
  Inf. Comp. Sci.} \textbf{1999,} \textsl{39,} 738-746.

\bibitem{Reynolds}
Reynolds,~C.~H. \textit{J. Comb. Chem.} \textbf{1999,} \textsl{1,} 297-306.

\bibitem{Linusson}
Linusson,~A.;\ \ Wold,~S.;\ \ Nord\'{e}n,~B. \textit{Mol. Divers.}
  \textbf{1999,} \textsl{4,} 103-114.

\bibitem{Deem}
Falcioni,~M.;\ \ Deem,~M.~W. \textit{Phys. Rev. E} \textbf{2000,} \textsl{61,}
  5948-5952.

\bibitem{Gobbi}
Gobbi,~A.;\ \ Poppinger,~D. \textit{Biotechnol. Bioeng.} \textbf{1998,}
  \textsl{61,} 47-54.

\bibitem{Weber}
Weber,~L.;\ \ Wallbaum,~S.;\ \ Broger,~C.;\ \ Gubernator,~K. \textit{Angew.
  Chem. Int. Ed. Engl.} \textbf{1995,} \textsl{34,} 2280-2282.

\bibitem{Sheridan}
Sheridan,~R.~P.;\ \ Kearsley,~S.~K. \textit{J. Chem. Inf. Comp. Sci.}
  \textbf{1995,} \textsl{35,} 310-320.

\bibitem{Singh}
Singh,~J.;\ \ Ator,~M.~A.;\ \ Jaeger,~E.~P.;\ \ Allen,~M.~P.;\ \
  Whipple,~D.~A.;\ \ Soloweij,~J.~E.;\ \ Chowdhary,~S.;\ \ Treasurywala,~A.~M.
  \textit{J. Am. Chem. Soc.} \textbf{1996,} \textsl{118,} 1669-1676.

\bibitem{Brennan}
Brennan,~M.~B. \textit{Chem. Eng. News} \textbf{2000,} \textsl{78,} 63-73.

\bibitem{Brown1997}
Brown,~R.~D.;\ \ Clark,~D.~E. \textit{Expert Opinion on Therapeutic Patents}
  \textbf{1997,} \textsl{8,} 1447-1459.

\bibitem{Brown1998}
Brown,~R.~D.;\ \ Martin,~Y.~C. \textit{J. Med. Chem.} \textbf{1998,}
  \textsl{40,} 2304-2313.

\bibitem{Katritzky}
Katritzky,~A.~R.;\ \ Kiely,~J.~S.;\ \ H\'{e}bert,~N.;\ \ Chassaing,~C.
  \textit{J. Comb. Chem.} \textbf{2000,} \textsl{2,} 2-5.

\bibitem{Kauffman}
Kauffman,~S.;\ \ Levin,~S. \textit{J. Theor. Biol.} \textbf{1987,}
  \textsl{128,} 11-45.

\bibitem{Perelson}
Perelson,~A.~S.;\ \ Macken,~C.~A. \textit{Proc. Natl. Acad. Sci. USA}
  \textbf{1995,} \textsl{92,} 9657-9661.

\bibitem{Deem2}
Bogard,~L.~D.;\ \ Deem,~M.~W. \textit{Proc. Natl. Acad. Sci. USA}
  \textbf{1999,} \textsl{96,} 2591-2595.

\bibitem{Frenkel}
Frenkel,~D.;\ \ Smit,~B. \textit{Understanding Molecular Simulation;} Academic
  Press: San Diego, 1996.

\bibitem{Fesik}
Shuker,~S.~B.;\ \ Hajduk,~P.~J.;\ \ Meadows,~R.~P.;\ \ Fesik,~S.~W.
  \textit{Science} \textbf{1996,} \textsl{274,} 1531-1534.

\bibitem{Moore}
Fejzo,~J.;\ \ Lepre,~C.~A.;\ \ Peng,~J.~W.;\ \ Bemis,~G.~W.;\ \ Ajay,;\ \
  Murcko,~M.~A.;\ \ Moore,~J.~M. \textit{Chem. Biol.} \textbf{1999,}
  \textsl{6,} 755-769.

\bibitem{Ellman}
Maly,~D.~J.;\ \ Choong,~I.~C.;\ \ Ellman,~J.~A. \textit{Proc. Natl. Acad. Sci.
  USA} \textbf{2000,} \textsl{97,} 2419-2424.

\bibitem{Griffey}
Griffey,~R.~H.;\ \ Hofstadler,~S.~A.;\ \ Sannes-Lowery,~K.~A.;\ \
  Ecker,~D.~J.;\ \ Crooke,~S.~T. \textit{Proc. Natl. Acad. Sci. USA}
  \textbf{1999,} \textsl{96,} 10129-10133.

\bibitem{Swendsen}
Swendsen,~R.~H.;\ \ Wang,~J.-S. \textit{Phys. Rev. Lett.} \textbf{1986,}
  \textsl{57,} 2607-2609.

\bibitem{Geyer}
Geyer,~C.~J.  Markov Chain {Monte} {Carlo} Maximum Likelihood.   In
  \textit{Computing Science and Statistics: Proceedings of the 23rd Symposium
  on the Interface}; American Statistical Association: New {York}, 1991.

\end{thebibliography}
\clearpage

\newpage
\begin{table}
\label{parameter}
\centering
\begin{minipage}[htbp]{6.00in}
\caption{Optimal parameters used in simulations for the three random energy models.}
\begin{tabular}{ccccccccccc} \hline
Model&$\beta$&$p_{\rm core}$&$p_{\rm swap}$&$p_{\rm swap_C}$&
$p_{\rm swap_S}$& $\beta_1$& $\beta_2$& $\beta_3$& $p_{\rm ex}$ \\
\hline I & 30 & 0.02 & 0.1 & 0.05 & 0.2 & 5 & 30 & 200 & 0.1 \\ \hline
II & 30 & 0.02 & 0.2 & 0.2 & 0.3 & 5 & 30 & 500 & 0.1 \\ \hline 
III & 50 & 0.02 & 0.4 & 0.2 & 0.2 & 5 & 50 & 500 & 0.1 \\ \hline
\end{tabular}
\end{minipage}
\end{table}

\clearpage
\newpage

\newpage

\newpage

\newpage

\end{document}